\documentclass[
reprint,
superscriptaddress,
amsmath,
amssymb,
aps,
longbibliography,
pra,
showpacs,
floatfix
]{revtex4-2}

\usepackage{hyperref}
\usepackage{graphicx}
\usepackage{xcolor} 
\usepackage{amsmath} 
\usepackage{booktabs}
\usepackage[rightcaption]{sidecap}
\usepackage{natbib}
\usepackage{makecell}

\usepackage{tabularx}
\usepackage{textcomp} 

\usepackage{multirow}
\usepackage{gensymb}

\usepackage{comment}
\usepackage{enumitem}
\usepackage{capt-of} 

\usepackage[utf8]{inputenc}

\hypersetup{colorlinks=true, 
    linkcolor={blue},
    citecolor={blue}, 
    urlcolor={blue}
}

\begin{document}
\makeatletter
\def\@parskip{0pt}
\def\@parindent{15pt}
\makeatother

\title{Nonlinear phononic slidetronics}

\author{Pooja Rani}
\email{p.rani@tue.nl}
\affiliation{School of Physics and Astronomy, Tel Aviv University, Tel Aviv 6997801, Israel}
\affiliation{Department of Applied Physics and Science Education, Eindhoven University of Technology, 5612 AP Eindhoven, Netherlands}

\author{Dominik~M.\ Juraschek}
\email{d.m.juraschek@tue.nl}

\affiliation{School of Physics and Astronomy, Tel Aviv University, Tel Aviv 6997801, Israel}
\affiliation{Department of Applied Physics and Science Education, Eindhoven University of Technology, 5612 AP Eindhoven, Netherlands}

\date{\today}


\begin{abstract}
Van der Waals ferroelectrics are conventionally switched by sliding the different layers between stacking orders with opposing electric polarizations. Ultrashort laser pulses have been proposed to launch shear modes and induce switching, with often unfeasible large pulse energies however. Here, we demonstrate switching of ferroelectricity in bilayer hexagonal boron nitride through nonlinearly excited phonons. We show that the efficiencies of conventional coherent phonon excitation mechanisms, including infrared absorption and Raman scattering techniques, are too low to overcome the energy barrier separating the two ferroelectric states. We demonstrate instead that excitation of high-frequency intralayer modes leads to a tilting of the interlayer potential-energy landscape that enables changing the stacking order. Our results provide an avenue towards efficient phononic slidetronics, enabling ultrafast control of the stacking order in van der Waals materials.
\end{abstract}

\maketitle


\section{Introduction}

Ultrafast control of ferroelectricity promises applications in advanced data processing devices operating on femto- and picosecond timescales. Recent advances have focused on the manipulation of the crystal structure itself through coherent and nonlinear excitation of lattice vibrations (phonons) with ultrashort laser pulses, which achieved transient switching \cite{Subedi2015,Mankowsky2017,Henstridge2022,chen2022deterministic,Kwaaitaal2024,shi2024nonresonant} and inducing \cite{Li2019,Nova2019,Shin2022,Fechner2024} of ferroelectricity. A major challenge in bulk ferroelectric materials is posed by the electric depolarizing field that prevents permanent reversal of the ferroelectric polarization in these excitation protocols \cite{Mankowsky2017,Abalmasov2020,Abalmasov2021}. Van der Waals ferroelectrics, in contrast, may circumvent this problem due to the reduced dimensionality of bi- and multilayer systems \cite{Wu2021_sliding,Wang2023}.

Van der Waals ferroelectrics can be switched by changing the stacking configuration when sliding the different layers with respect to each other, which has lead to the emergent field of sliding ferroelectricity \cite{vizner2021interfacial,Meng2022,Yasuda2024,Bian2024,ViznerStern2025,deng2025deterministic}. From a phononic perspective, the sliding motion can be expressed in terms of low-frequency shear modes that can be launched with ultrashort laser pulses \cite{Sie2019,Zhang2019,Fukuda2020,rostami2022}. Recent theoretical studies have since suggested that these light-driven shear-mode dynamics could be utilized to switch the stacking order and thus ferroelectric polarization in various van der Waals materials \cite{Park2019,rostami2022,Wang2024,Yang2024,Gao2024,wei2025ultrafast}. A major challenge is posed by the weak light-matter couplings that make ferroelectric switching unfeasible with experimentally achievable pulse energies.

Here, we demonstrate switching of the ferroelectric polarization in bilayer hexagonal boron nitride (h-BN) through nonlinearly driven phonons. Using a combination of first-principles calculations and phenomenological modeling, we show that all common mechanisms of coherent phonon excitation for large band gap semiconductors and insulators, including infrared absorption (IRA), ionic Raman scattering (IRS) \cite{Forst2011,Subedi2014}, impulsive stimulated Raman scattering (ISRS) \cite{merlin1997generating}, terahertz sum-frequency excitation (THz-SFE) \cite{Maehrlein2017,Juraschek2018}, and infrared resonant Raman scattering (IRRS) \cite{Khalsa2021}, do not yield sufficiently large shear-mode amplitudes to overcome the energy barrier between the two stacking orders. We then in contrast show how coherently excited intralayer modes tilt the interlayer potential-energy landscape and enable switching for experimentally achievable pulse energies. This involves a dynamical modification of the shear modes potential beyond conventional phononic rectification \cite{Subedi2014,fechner:2016,Juraschek2017,Khalsa2024}. Our calculations show that nonlinear phononic slidetronics provides efficient and ultrafast control over the stacking order in van der Waals materials.


\begin{figure*}
\centering
\includegraphics[width=16cm]{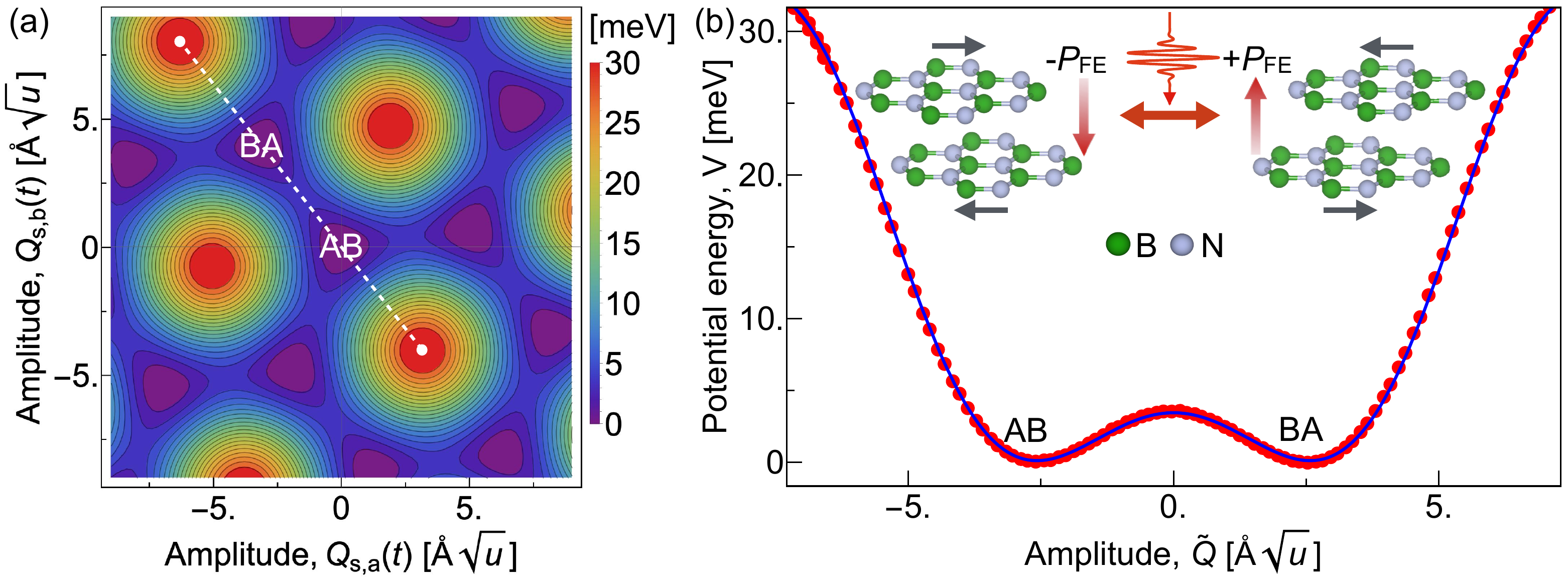}
\caption{Phononic sliding in hexagonal boron nitride. (a) Potential energy landscape as a function of  shear-mode amplitude. The six-fold rotationally symmetric minima correspond to AB and BA stacking orders with opposing ferroelectric polarizations, $P_\text{FE}$. The white dashed line corresponds to the double-well potential shown in (b). Double-well potential of the ferroelectric AB and BA stacking orders. In AB stacking, nitrogen atoms from the upper layer are centered over the hexagonal ring, whereas boron atoms from the upper layer are located over nitrogen atoms from the lower layer---vice versa for BA stacking. The red dots are calculated data and the blue line is a fit. Grey arrows denote motion of the layers along the shear mode.
}
\label{fig:Figure1}
\end{figure*}


\section{Properties of h-BN}

We begin by calculating the properties of h-BN from first principles using density functional theory. We computed phonon eigenfrequencies, eigenvectors, Raman tensors, and nonlinear phonon couplings. We obtain a large band gap of 6.3~eV, consistent with previous calculations~\cite{wickramaratne2018monolayer} and suitable for phonon driving using strong pulses. We find the eigenfrequency of the doubly degenerate shear mode with $E$ symmetry to be $\Omega_s/(2\pi)=0.94$~THz. For the full list of phonon modes and computational details, see Supplemental Material \cite{SUPP}. We show the calculated potential energy landscape formed by the two orthogonal shear-mode components, $Q_{s,a}$ and $Q_{s,b}$, in Fig.~\ref{fig:Figure1}(a). The minima correspond to AB and BA stacking orders with opposing ferroelectric polarizations, illustrated in Fig.~\ref{fig:Figure1}(b). There, we show the double-well potential corresponding to the dashed linecut in (a). 

The goal of this study is to find an efficient mechanism that allows us to switch between the two stacking orders by the means of shear-mode excitation. We therefore next compute and compare the efficiencies of the five most prominent coherent phonon excitation mechanisms.


\section{Shear-Mode Excitation Mechanisms}

We consider five excitation mechanisms for the shear modes, IRA, IRS, ISRS, THz-SFE, and IRRS. We do not consider above-bandgap excitation mechanisms such as displacive excitation of coherent phonons (DECP) that would require ultraviolet pulses and lead to additional electronic dissipative effects \cite{Yang2024,wei2025ultrafast}. The coherent phonon dynamics can be described by semiclassical equations of motion, for which we compute the parameters from first principles \cite{Subedi2014,fechner:2016,Juraschek2018,VonHoegen2018},
\begin{align}\label{eq:eom}
\ddot{\mathbf{Q}}_s+\kappa\dot{\mathbf{Q}}_s+\Omega^2\mathbf{Q}_s+\mathbf{F}(t)= 0.
\end{align}
Here, $\mathbf{Q}_s=(Q_{s,a},Q_{s,b},0)$ contains the amplitudes of the two orthogonal components of the shear mode. We set the damping to $\kappa_s= 0.1\times\Omega_s/(2\pi)$ for the shear mode, in agreement with typical shear-mode dampings found in h-BN~\cite{kang2021ultrafast}. $\mathbf{F}(t)=-\nabla_\mathbf{Q} V_\mathrm{int}$ is the driving force that is given by the interaction potential, $V_\text{int}$, containing both light-matter and nonlinear phonon couplings. We model the laser pulse with $\mathbf{E}(t, \phi, \theta)=\tilde{E}_{0}\exp{[t^{2}/(2(\tau/\sqrt{8\ln2})^{2})]}(\cos(\omega_{0}t)\cos(
\theta), \cos(\omega_{0}t+\phi) \sin(\theta),0)$, where $\tilde{E}_0=2E_0/(1+\sqrt{\epsilon_\text{2D}})$ is the reduced electric field of the pulse, accounting for electronic screening in the material from the in-plane dielectric function, which we compute to be $\epsilon_\text{2D}=4.77$. $\tau$ is the full width at half maximum pulse duration and $\omega_0$ is the center frequency. $\phi$ and $\theta$ control the polarization state and orientation of the pulse and their respective values for the different mechanisms are provided in Supplemental Material. We further list $E_0$, $\tau$, and $\omega_0$ to be used for the five mechanisms in Table~\ref{tab:Inputs}. We adjust pulse duration and peak electric field so that the total pulse energy $\propto E_0^2\tau$ remains constant.


\begin{table}[h]
    \centering
    \caption{Parameters for the comparison of the five shear-mode excitation mechanisms. $\omega_0$ is the center frequency, $\tau$ is  full width at half maximum pulse duration, and $E_0$ is the peak electric field of the ultrashort laser pulse. The pulse energy $\propto E_0^2\tau$ is kept constant for all cases. $|\mathbf{Q}_s|$ denotes the maximum achieved shear-mode amplitude. For IRS, ISRS, and IRRS, two pulses are used each.}  
    \begin{tabular}{llllll}
    \hline\hline
       Mechanisms            & IRA~~     & IRS~~~    &ISRS~     & THz-SFE~~   &IRRS   \\
    \hline
     $\omega_{0}/(2\pi)$ (THz)  & 1   &40.5  & 374  &0.5  &40.5 \\
     
     $\tau$ (ps)               &  1   & 0.05  & 0.05  &2 &0.05 \\ 
     
      $E_{0}$ (MV/cm)           & 12.7 &40.2   & 40.2   & 9  &40.2\\
     
     $|\mathbf{Q}_s|$ (pm$\sqrt{u}$)  & 11   &8  &9.4$\times10^{-2}$ &4$\times10^{-2}$    &14$\times10^{-3}$  \\
    \hline\hline
    \end{tabular}
    \label{tab:Inputs}
\end{table}


\begin{figure*}[t]
\includegraphics[width=17.8cm]{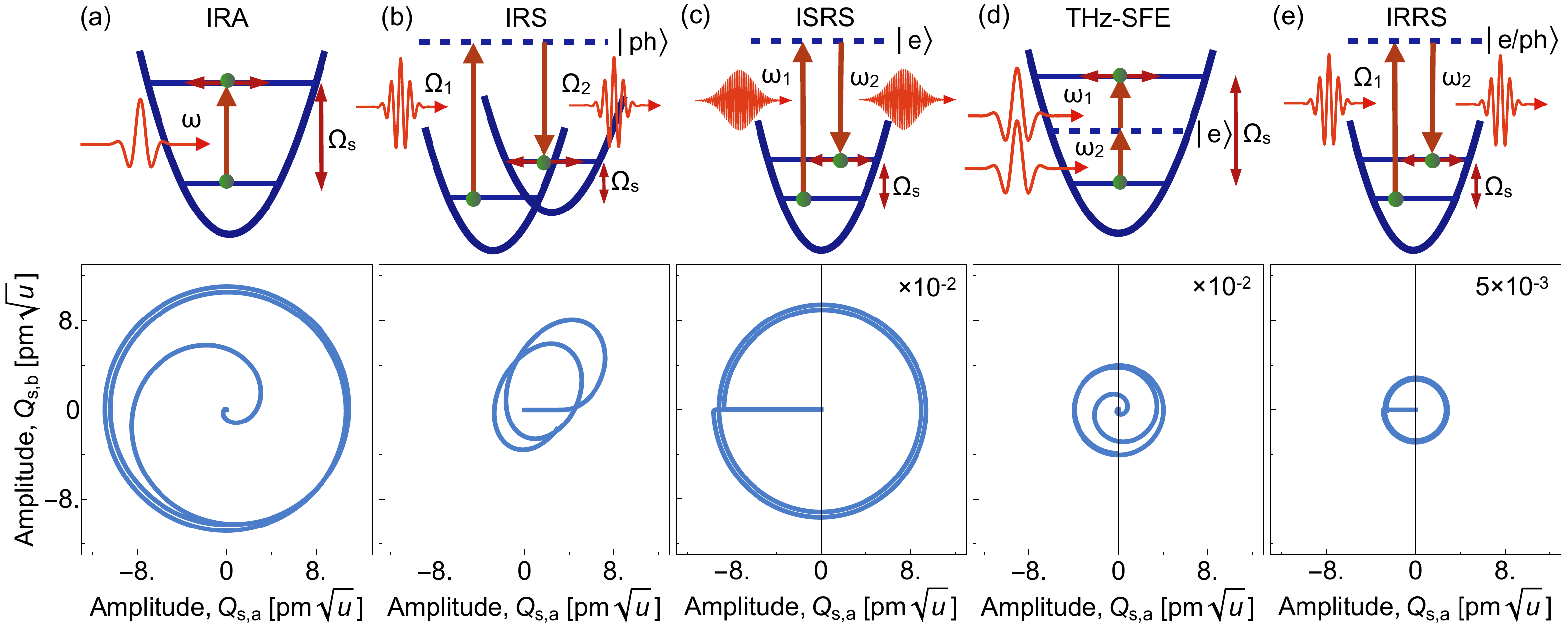}
\caption{Shear-mode excitation mechanisms for circular polarization. Energy diagrams (top row) depict the scattering processes, where red pulses indicate incoming and outgoing photons and dashed lines mark intermediate electronic $|\text{e}\rangle$ and phononic $|\text{ph}\rangle$ states in the Raman-type processes. We plot the time evolutions of the amplitudes of the two orthogonal shear-mode components following the excitation for a time window up to 6~ps. (a) Infrared absorption (IRA), (b) ionic Raman scattering (IRS), (c) impulsive stimulated Raman scattering (ISRS), (d) terahertz sum-frequency excitation (THz-SFE), and (e) infrared resonant Raman scattering (IRRS).
}
 \label{fig:Figure2}
  \end{figure*}

  
\subsection{Infrared absorption}

The first mechanism we investigate is IRA, which is described by the direct coupling of the electric dipole moment of the phonon mode to the electric field of light, $V_\text{int} = Q_{s,a/b}\mathbf{Z}\cdot\mathbf{E}(t)$. Here, $\mathbf{Z}=\sum_\alpha Z_\alpha^* \mathbf{q}_{s,\alpha}/\sqrt{M_\alpha}$ is the mode effective charge vector, where $Z^*$ is the Born effective charge tensor, $\mathbf{q}_{s,\alpha}$ is the phonon eigenvector, and $M_\alpha$ the atomic mass of atom $\alpha$ in the unit cell. This mechanism requires resonant excitation of the phonon mode, $\omega_0=\Omega_s$, and corresponds to a one photon absorption process by the phonon, illustrated in Fig.~\ref{fig:Figure2}(a). In semiconductors and insulators, IRA is generally the most efficient mechanism to induce coherent phonon oscillations with large amplitudes, leading to the field of nonlinear phononics within the past decade \cite{Forst2011,Subedi2014,Disa2021}. While in bulk oxides, $|\mathbf{Z}|\sim1~e/\sqrt{u}$ \cite{Juraschek2019,Pueyo2022,Khalsa2024}, where $e$ is the elementary charge and $u$ the atomic mass unit, we find the mode effective charge of the shear modes in h-BN to be only $|\mathbf{Z}_s|=0.002~e/\sqrt{u}$. The dynamics according to Eq.~\eqref{eq:eom} yield a maximum phonon amplitude of $|\mathbf{Q}_s|$ = 11~pm$\sqrt{u}$, which we show in Fig.~\ref{fig:Figure2}(a). This weak light-matter coupling explains why giant pulse energies were needed to predict switching of a van der Waals ferroelectric with infrared absorption in a previous study~\cite{Wang2024}.


\subsection{Ionic Raman scattering}

The second mechanism we investigate is IRS. Because of the lack of inversion symmetry, the shear modes are both infrared and Raman active. In IRS, the laser pulse first excites high-frequency intralayer phonon modes through infrared absorption that then couple to the shear modes through nonlinear phonon interactions \cite{Forst2011,Subedi2014,mankowsky2016non}. At the same time, the phonon excitation leads to rectification of the coupled phonon mode, also called phononic rectification, see Fig.~\ref{fig:Figure2}(b). In bulk ferroelectrics, phononic rectification has been previously used to transiently switch ferroelectricity \cite{Subedi2015,mankowsky2016non,Abalmasov2020,Abalmasov2021,Henstridge2022}. The interaction potential is given by $V_\text{int}(t)= 2c_n Q_{n,x}Q_{n,y}Q_{s,a}+c_n (Q_{n,x}^{2}-Q_{n,y}^{2})Q_{s,b} + Q_{n,x/y}\mathbf{Z}_n\cdot\mathbf{E}(t)$, where $\mathbf{Q}_n=(Q_{n,x},Q_{n,y},0)$ is the amplitude vector of a high-frequency intralayer mode, $n$, $\mathbf{Z}_n$ is its mode effective charge vector, and $c_n$ is the nonlinear phonon coupling. The calculated eigenfrequencies of the two high-frequency $E$ modes are $\Omega_1/(2\pi)=40.45$~THz and $\Omega_2/(2\pi)=40.5$~THz. We find the mode effective charges to be $|\mathbf{Z}_1|=1.4~e/\sqrt{u}$ and $|\mathbf{Z}_2|=0.7~e/\sqrt{u}$, as well as the nonlinear phonon couplings to be $c_1=0.93$~meV/(\AA$\sqrt{u}$$)$$^3$ and $c_2=5.3~$meV/(\AA$\sqrt{u}$$)$$^3$, respectively. The damping for high-frequency mode we took is $\kappa_n= 0.02\times\Omega_n/(2\pi)$ which is in good agreement with the literature~\cite{Juraschek2021, giles2018ultralow, cusco2018isotopic}. To evaluate the phonon dynamics, Eq.~\eqref{eq:eom} is expanded to include equations of motion for the high-frequency phonon modes, see Supplemental Material for details~\cite{SUPP}. In order to achieve circular excitation, two linearly polarized laser pulses have to be applied with a time delay of $\Delta t = 1/(4\Omega_s)$ and an angle of $\theta=45^\circ$. Both pulses are tuned to resonance with the high-frequency phonon modes, $\omega_0/(2\pi)=40.5$~THz. We show the dynamics in Fig.~\ref{fig:Figure2}(b). The shear-mode amplitude generated by IRS reaches $|\mathbf{Q}_s|$ = 8~pm$\sqrt{u}$, which is slightly smaller than that arising from infrared absorption. The weak response is a result of the small cubic-order nonlinear phonon couplings. These factors are the reason why giant pulse energies were needed to predict switching of a van der Waals ferroelectric through conventional nonlinear phononics in a previous study \cite{Park2019}.


\subsection{Impulsive stimulated Raman scattering and Terahertz sum-frequency excitation }

The third and fourth mechanisms we investigate are impulsive ISRS and THz-SFE that are described by the same fundamental coupling term. In ISRS, the difference frequency of two photons from the laser pulse is resonant with the shear mode, as illustrated in Fig.~\ref{fig:Figure2}(c)~\cite{merlin1997generating}, while in THz-SFE, two photons from the ultrashort laser pulse are absorbed by a phonon, as illustrated in Fig.~\ref{fig:Figure2}(d) \cite{Maehrlein2017,Juraschek2018,Melnikov2018,Knighton2019,Johnson2019,Melnikov2020,Blank2023_SF-IRS,Kusaba2024,Levchuk2025,Minakova2025}. The interaction potential for both mechanisms is given by $V_\text{int}(t)=2R\epsilon_{0}E_xE_yQ_{s,a}+R\epsilon_{0}(E_x^2-E_y^2)Q_{s,b}$, where $R$ is the single independent component of the Raman tensor. Our calculations yield $R(1.55~\text{eV})=0.09$~\AA$^2\sqrt{u}$ and $R(0.47~\text{eV})=0.04$~\AA$^2\sqrt{u}$, see Supplemental Material for the full frequency dependence and symmetry~\cite{SUPP}. As in IRS, in order to achieve circular excitation for ISRS, two linearly polarized pulses have to be applied with a time delay of $\Delta t = 1/(4\Omega_s)$ and an angle of $\theta=45^\circ$. For THz-SFE in turn, a circularly polarized pulse is used. We show the shear-mode dynamics in Fig.~\ref{fig:Figure2}(c-d). The maximum amplitude reaches $|\mathbf{Q}_s| = 9.4\times 10^{-2}$~pm$\sqrt{u}$ for ISRS and $4\times 10^{-2}$~pm$\sqrt{u}$ for THz-SFE. The amplitude for both mechanisms are much smaller than those of IRA and IRS, because of the particularly small values of the Raman tensor (three orders of magnitude smaller than that of diamond \cite{Juraschek2018}). 


\subsection{Infrared resonant Raman scattering}

The last mechanism we investigate is infrared resonant raman scattering (IRRS), in which the difference frequency of the photon and phonon is resonant with the shear modes~\cite{khalsa2021ultrafast}, as shown in Fig.~\ref{fig:Figure2}(e). The interaction potential $V_\text{int}(t)= Q_{n.x/y}(b\mathbf{E}(t)Q_{s}+ \mathbf{Z}\cdot \mathbf{E}(t))$, where the coefficient $b$ is defined as $\partial \mathbf{Z}_{n}/\partial Q_{s}$. The electric field $\mathbf{E}(t)$ appearing in the term $\mathbf{Z}\cdot \mathbf{E}(t)$ corresponds to the field used to excite high-frequency intralayer modes in the IRS mechanism, while $\mathbf{E}(t)$ appearing in the term $b \mathbf{E}(t) Q_{s}$ represents the field component responsible for driving the shear modes. To excite the shear modes, we used two linearly polarized pulses with a time delay of $\Delta t = 1/(4\Omega_s)$ and an angle of $\theta=-45^\circ$. The values of $b$ we calculated for $\mathbf{Q}_{1}$ and $\mathbf{Q}_{2}$ are $0.001~e/(\text{\AA}u)$ and $0.0025~e/(\text{\AA}u)$, respectively . The shear-mode dynamics is shown in Fig.~\ref{fig:Figure2}(e), which produces a maximum amplitude of $|\mathbf{Q}_{s}| = 14\times 10^{-3}$~pm$\sqrt{u}$, which is lower than the four above mechanisms. 

Our simulations of the shear mode amplitudes in this section produce phonon amplitudes of up to 11~pm$\sqrt{u}$ for IRA and 8~pm$\sqrt{u}$ for IRS, whereas the other mechanisms yield much smaller amplitudes. It shall be noted that the phonon amplitude generated by IRA scales linearly with the electric field, whereas all Raman-type mechanisms scale quadratically, possibly leading to larger amplitudes as the pulse energy increases. To overcome the energy barrier however, an amplitude of at least 2.58~\AA$\sqrt{u}$ is required. We therefore evaluated the critical pulse energies at which the different mechanisms would lead to ferroelectric switching, see Supplemental Material \cite{SUPP}. All five mechanisms require unfeasibly strong pulses, making switching based on conventional coherent excitation impractical. We therefore investigate an alternative approach that efficiently reduces the switching threshold, by explicitly taking into account the time-dependent double-well potential of the shear modes upon strong driving of the high-frequency intralayer modes. 


\section{Ferroelectric switching with nonlinear phononic slidetronics}

\begin{figure}[t]
\centering
\includegraphics[width=8cm]{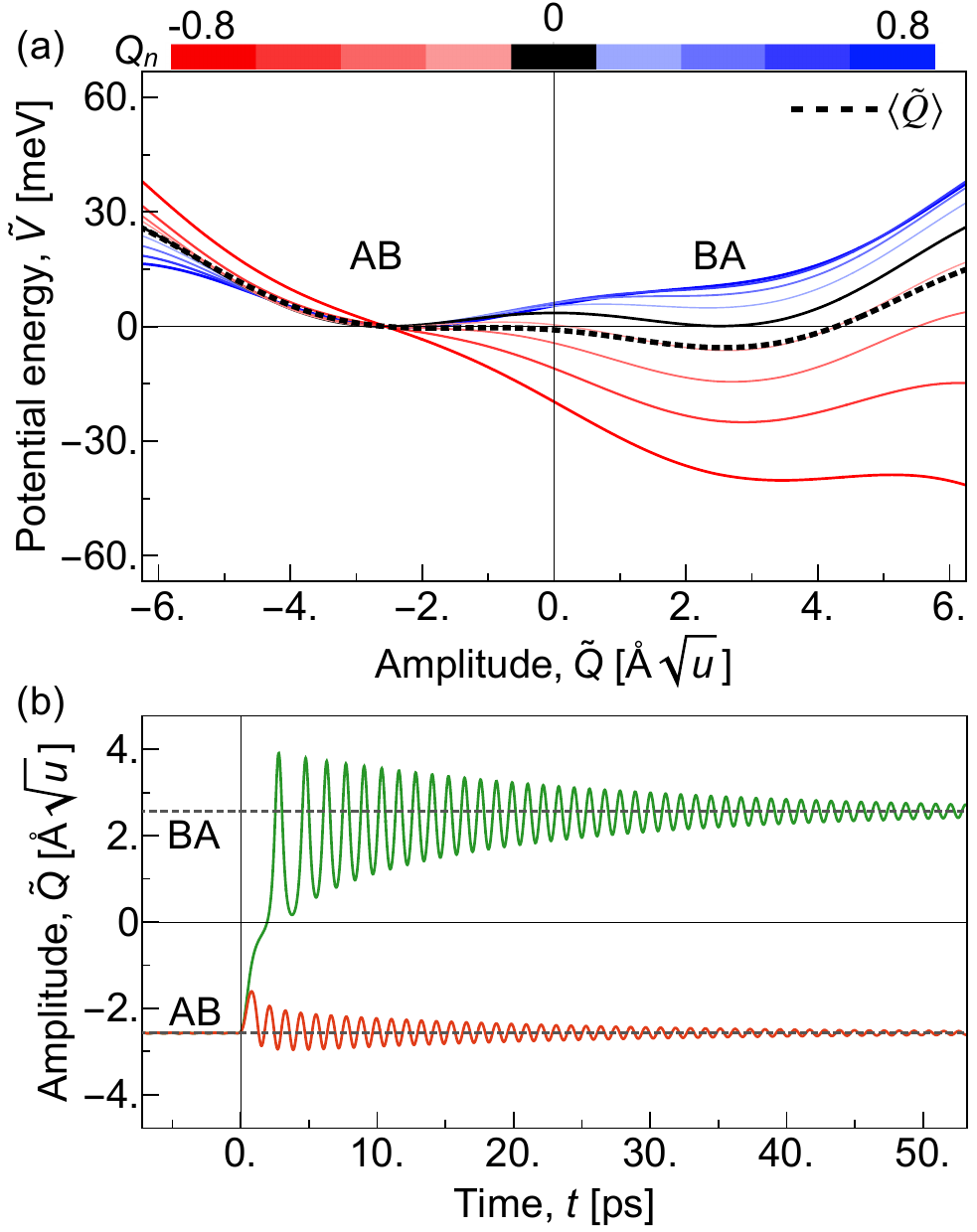}
\caption{Ferroelectric switching through nonlinear phononic slidetronics. (a) Tilting of the double-well potential upon displacement of the high-frequency intralayer $E$ mode at 40.5~THz. The dotted black curve is the average of $\tilde{V}(\tilde{Q})$ obtained from tiltings due to the maximum and minimum amplitudes of the 40.5~THz mode. (b) Time evolution of shear-mode amplitudes following the excitation of the 40.5~THz mode. The red curve shows rectified coherent oscillations of the shear mode for a pulse energy below the switching threshold ($E_0$ = 17.5~MV/cm), whereas the green curve shows a changing of the stacking order from AB to BA for a pulse energy above the switching threshold ($E_0$ = 20~MV/cm).
}
\label{fig:Figure3}
\end{figure}

The shear modes must be launched along the path indicated in Fig.~\ref{fig:Figure1}(a), parametrized by the generalized shear-mode coordinate, $\tilde{Q}$, in order to transition between AB and BA stacking orders. The potential energy of the double well can be written as $\tilde{V} = \alpha \tilde{Q}^{2}+\beta \tilde{Q}^{4}+\gamma \tilde{Q}^{6}+\delta\tilde{Q}^{8}$. The equation of motion for $\tilde{Q}$ accordingly becomes
\begin{align}
\ddot{\tilde{Q}}+\kappa \dot{\tilde{Q}}+\frac{\partial \tilde{V}}{\partial \tilde{Q}}= 0. 
\end{align}
In Fig.~\ref{fig:Figure3}(a), we show calculations for the tilting of the potential-energy landscape of the shear modes upon displacing the $E$ mode at 40.5~THz, $Q_2$, leading to a dynamical lifting of the degeneracy of the energy minima of AB and BA stacking orders. While the potential will tilt back and forth at high frequency for positive and negative amplitudes of $Q_2$, the dynamical average $\tilde{Q}$ still shows a net tilting. The $E$ mode at 40.45~THz only leads to weak modulations and is therefore neglected here. As a result of the tilting, the double-well potential gains contributions with odd powers in the coordinate $\tilde{Q}$, and we can model the coefficients of the double-well potential as time-dependent functionals of the 40.5~THz $E$-mode amplitude, $\tilde{V}[Q_2(t)]$, see Supplemental Material for the full expression~\cite{SUPP}. 

In Fig.~\ref{fig:Figure3}(b), we show the temporal evolution of the generalized shear-mode coordinate following the excitation. The 40.5~THz mode starts ringing and dynamically tilts the double-well potential. We find the switching threshold to be reached when the 40.5~THz mode is excited using a laser pulse with a peak electric field of $E_0=20$~MV/cm and a pulse duration of $\tau = 0.5$~ps. The orientation of the pulse is described by $\phi=\Delta t=\theta= 0$. These specifications lie well within experimentally achievable pulse strengths \cite{liu2017,Vicario2020}. The generalized shear-mode coordinate crosses the energy barrier starting from AB stacking and then relaxes into the BA stacking configuration, reversing the ferroelectric polarization in the process. For an excitation with a pulse energy below the switching threshold, here $E=17.5$~MV/cm, the shear mode simply gets coherently excited and relaxes back into the original stacking configuration.


\section{Discussion}

Our results demonstrate that ferroelectricity in bilayer h-BN can be efficiently switched by tilting the interlayer shear-mode potential through high-frequency intralayer phonon excitation. The pulse strengths required to overcome the energy barrier between stacking orders are readily achievable with state-of-the-art tabletop pump-probe setups and the reversal of ferroelectric polarization could be measured with terahertz pump-second harmonic generation probe measurements. The mechanism we describe is generally applicable to bi- and multilayer van der Waals materials, allowing for light-induced control over stacking order that can be extended to other electronic orders, including magnetism and topology \cite{Sie2019,rodriguez2020phonon,rodriguez2021direct,lin2025interlayer}.

We note that the switching threshold is sensitive to the damping of the shear modes that allows them to deterministically relax into one of the energy minima and therefore stacking configuration. Phonon damping in van der Waals materials can be strongly influenced by sample quality and even isotopic purity~\cite{giles2018ultralow, cusco2018isotopic}. For large damping and short phonon lifetimes, the pulse energy to reach the switching threshold increases. In turn, for small damping and long phonon lifetimes, the shear mode can oscillate back and forth between the two stacking orders which requires fine tuning of the pulse strength. We show corresponding simulations in Supplemental Material~\cite{SUPP}. We use shear-mode dampings matching those found in experiment \cite{kang2021ultrafast}, however these strictly only apply to displacements of the shear-mode coordinate around the equilibrium position. Not captured by our model are explicit couplings of the shear modes to the acoustic branches \cite{teitelbaum2018direct,Khalsa2024}, which may be modified out of equilibrium and lead to modifications of the damping during the dynamics. A more comprehensive treatment therefore will need to take into account amplitude- and time-dependent damping modulations in the future.

Our results demonstrate nonlinear phononic slidetronics as a viable avenue for changing the stacking order and therefore structural and electronically ordered phases of van der Waals materials. This mechanism could serve as a base for highly efficient all-optically controlled low-dimensional ferroelectric memory devices in the future.
  

\begin{acknowledgments}
We thank E. Baldini, D. Kaplan, L. Benfatto, and E. Demler for useful discussions. This work was supported by the Army Research Office grant No. W911NF-23-1-0243. Calculations were performed on HPC infrastructure at Tel Aviv University.
\end{acknowledgments}



%


\onecolumngrid
\clearpage

\setcounter{page}{1}

\begin{center}
\textbf{\large Supplemental Material: Nonlinear phononic slidetronics}\\[0.4cm]
Pooja Rani,$^{1}$,$^{2}$ , Dominik~M.~Juraschek$^{1}$,$^{2}$\\[0.15cm]
\affiliation{}
$^1${\itshape{\small School of Physics and Astronomy, Tel Aviv University, Tel Aviv 6997801, Israel}}\\
$^2${\itshape{\small Department of Applied Physics and Science Education, Eindhoven University of Technology, 5612 AP Eindhoven, Netherlands}}\\
\end{center}

\date{\today}

\setcounter{equation}{0}
\setcounter{figure}{0}
\setcounter{table}{0}
\makeatletter
\renewcommand{\theequation}{S\arabic{equation}}
\renewcommand{\thefigure}{S\arabic{figure}}
\renewcommand{\thetable}{S\arabic{table}}


\section*{Computational details}

We calculated the properties of h-BN from first principles using the density functional theory formalism implemented in VASP \cite{Kresse1996, Kresse1996a} and the frozen phonon approach implemented in phonopy \cite{Togo2015}. We computed phonon eigenfrequencies, eigenvectors, Raman tensors, and nonlinear phonon couplings. We used the Perdew-Burke-Ernzerhof (PBE) exchange-correlation functional along with the generalized gradient approximation (GGA) \cite{Csonka2009}. Plane waves were set to a cutoff energy of 600~eV. The Brillouin zone was sampled using a Monkhorst-Pack $k$-point mesh \cite{Pack1977} of 11$\times$11$\times$1 for structural relaxation and phonon calculations. We converged the atomic forces to $10^{-4}$~eV/\AA{} accuracy. To model the bilayer structure with four atoms per unit cell, we included a vacuum of 20~\AA{}. The optimized bond length of 1.45~\AA{} and the interlayer distance of 3.34~\AA{} are in good agreement with typical literature values \cite{Constantinescu2013, Lebedev2016}.


\section*{Phonon frequencies and eigenvectors}
\begin{figure*}[h]
\centering
\includegraphics[width=14cm]{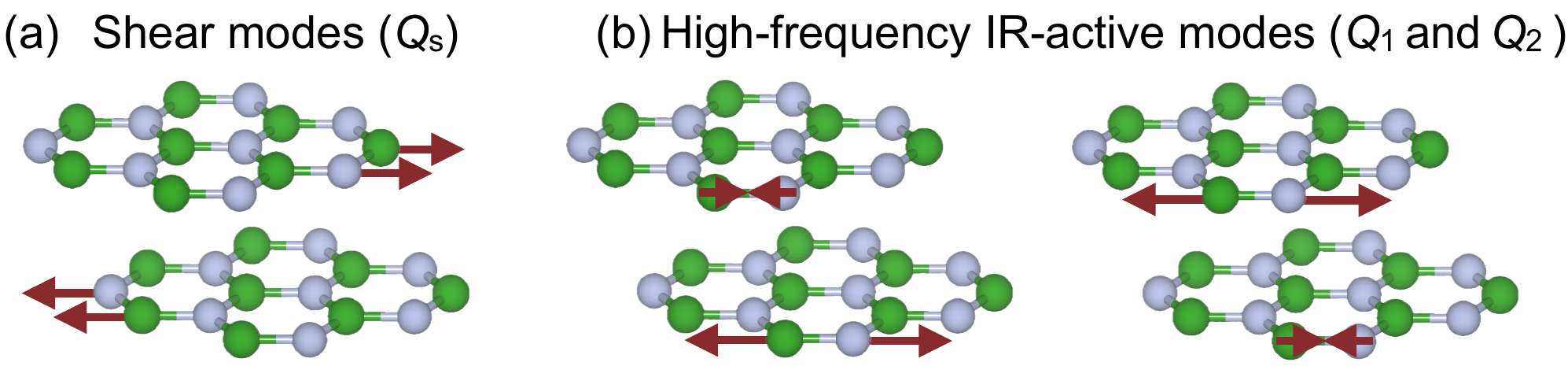}
\caption{Visualization of eigenvectors. (a) Shear modes showing relative motion of the layers. (b) High-frequency $E$ modes showing the intralayer motion of the atoms, where the lengths of the arrows indicate the relative atomic displacements.
}
\label{fig:eigenvectors}
\end{figure*}

Bilayer h-BN in AB stacking order has a point group of $C_{3v}$ and exhibits 12 zone-center phonon modes that can be described by the irreducible representations $4E\oplus4A_{1}$, of which 9 represent optical phonon modes, as shown in Table~\ref{tab:phonon_modes}. Here, the lowest-energy degenerate optical modes are the shear modes, which move the different layers with respect to each other and are both IR and Raman active. Modes 6, 7, and 8 are out-of-plane $A_1$ modes, where mode 6 corresponds to the breathing mode. The high-frequency degenerate $E$ modes (9, 10, 11, and 12) are IR active and can be directly excited by the laser pulse through infrared absorption. Fig.~\ref{fig:eigenvectors} shows the eigenvectors of the shear and high-frequency IR-active modes. The shear modes induce relative motion between the layers, enabling interlayer sliding. In contrast, the high-frequency IR-active modes are intralayer vibrations, which occur when atoms move relative to each other within the same layer without moving the layers themselves.


\begin{table}[h]
    \centering
    \caption{Calculated phonon modes at the $\Gamma$ point and their irreducible representations (I.R.) of AB-stacked bilayer h-BN.}
    \begin{tabular}{llll}
    \hline\hline
     No.~~~   &  Frequencies in THz ($\text{cm}^{-1}$)  &  I.R. \\
    \hline
     4,5   & 0.94 (31.3)     & $E$     \\ 
     6     & 2.3  (76.8)     & $A_{1}$  \\
     7     & 23.89 (796.8)   & $A_{1}$  \\
     8     & 23.98 (799.8)   & $A_{1}$  \\
     9,10  & 40.45 (1349.1)  & $E$       \\
     11,12 & 40.5  (1350.6)  & $E$       \\     
    \hline\hline
    \end{tabular}
    \label{tab:phonon_modes}
\end{table}


\section*{Laser pulse specifications}
For all mechanisms, the laser pulse is modeled in the form of a time-dependent Gaussian envelope electric field $\mathbf{E}(t)$, which we express as
 \begin{equation}
   \textbf{E}(t, \phi, \theta) = \tilde{E}_{0}\exp{\left[\frac{t^2}{2\big(\frac{\tau}{\sqrt{8\ln2}}\big)^{2}}\right]}
    \begin{pmatrix}
       \cos(\omega_{0} t)\cos(\theta)\\
       \cos(\omega_{0}t+ \phi)\sin(\theta)
   \end{pmatrix},
 \end{equation}
where $\tilde{E}_{0}=2E_0/(1+\sqrt{\epsilon_{2D}})$ is the reduced electric field of the pulse, which includes electronic
screening in the material from the in-plane dielectric
function $\epsilon_{2D}$= 4.77 calculated using first-principle calculations and consistent with the literature \cite{Gilbert2019,Laturia2018,Ohba2001}. $E_{0}$ is peak electric field and $\tau$, $\omega_0$ are the pulse duration and pulse frequency, respectively. 


\begin{table}[h]
    \centering
    \caption{Pulse parameters for the laser pulses used in the five excitation mechanisms.}
    \begin{tabular}{lllllll}

    \hline\hline
       Mechanisms    &No. of pulse ~~~  & IRA ~~  &IRS     &ISRS    &  THz-SFE &IRRS\\
    \hline

     $\phi$  & First pulse      & -$\pi/{2}$     & $\pi$    &0       & $\pi/{2}$  &0\\  
             & Second pulse     & -     & $\pi$   &0        & -  &0\\
\hline
   $\Delta t$  & First pulse    & 0      & 0     & 0                               & 0   &0\\
             & Second pulse      & -      &$1/(4\Omega_{s})$  &$1/(4\Omega_{s})$   & -    & $1/(4\Omega_{s})$  \\
\hline 
   $\theta$  & First pulse      & $\pi/{4}$     & $\pi/{4}$ 
                & $\pi/{4}$   &$\pi/{4}$     &0\\   
             & Second pulse      & -   & $\pi/{2}$   &-$\pi/{2}$            & -    &-$\pi/{4}$\\            
    \hline\hline
    \end{tabular}
    \label{tab:Values}
\end{table}


\section*{Raman-tensor calculations}


\subsection*{Ionic Raman scattering}

\begin{figure}[b]
\centering
\includegraphics[width=17cm]{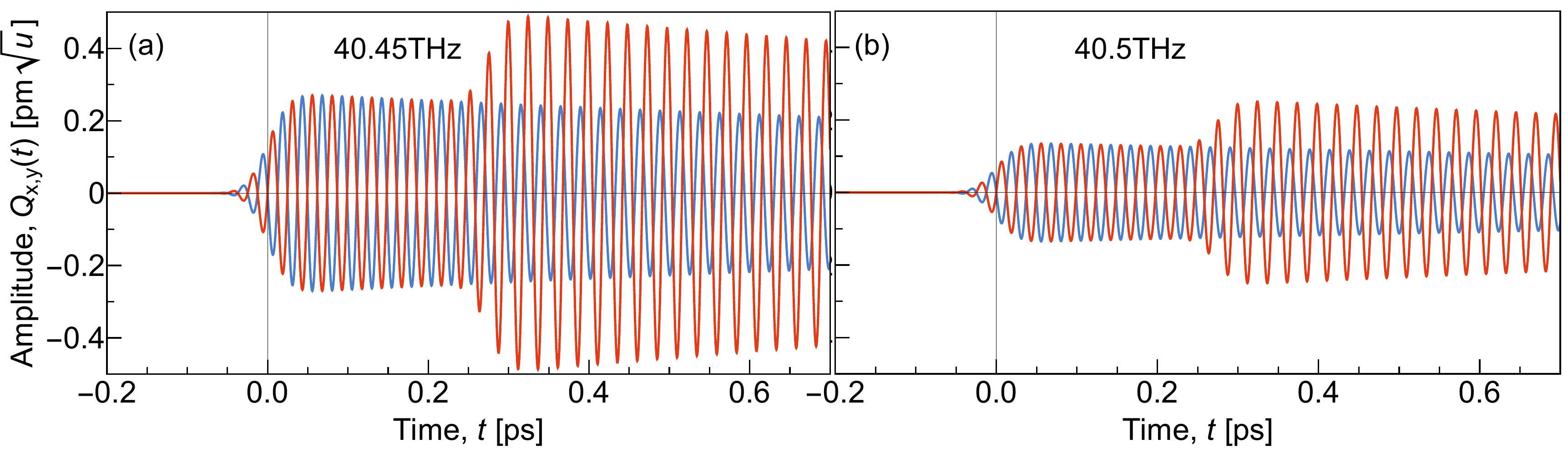}
\caption{Time evolution of the amplitudes of the high-frequency intralayer modes following the excitation by an ultrashort pulse with a peak electric field of $E_0= 40.2$~MV/cm and a pulse duration of $\tau=0.05$~ps. Blue graphs represent the amplitudes for $Q_{n,x}$, and red graphs represent the amplitudes for $Q_{n,y}$. We show plots for the (a) 40.45~THz mode and (b) 40.5~THz mode.
}
\label{fig:modeIR}
\end{figure}

In IRS, resonantly driven high-frequency intralayer phonon modes couple nonlinearly to the shear modes. The equation of motion for these intralayer modes reads
\begin{align}\label{eq:eomIR}
\ddot{\mathbf{Q}}_n+\kappa\dot{\mathbf{Q}}_n+\Omega^2\mathbf{Q}_n+\mathbf{Z}\cdot\mathbf{E}(t)= 0,
\end{align}
Where $\mathbf{Q}_n=(Q_{n,x},Q_{n,y},0)$ contains the two orthogonal components of the high-frequency intralayer modes. The indices $n= 1,2$ label the modes at 40.45~THz and 40.5~THz, respectively. The phonon dynamics following the excitation by a resonant laser pulse are shown in Fig.~\ref{fig:modeIR} for both $\mathbf{Q}_1$ and $\mathbf{Q}_2$.


\begin{figure}[h]
\centering
\includegraphics[width=17.5cm]{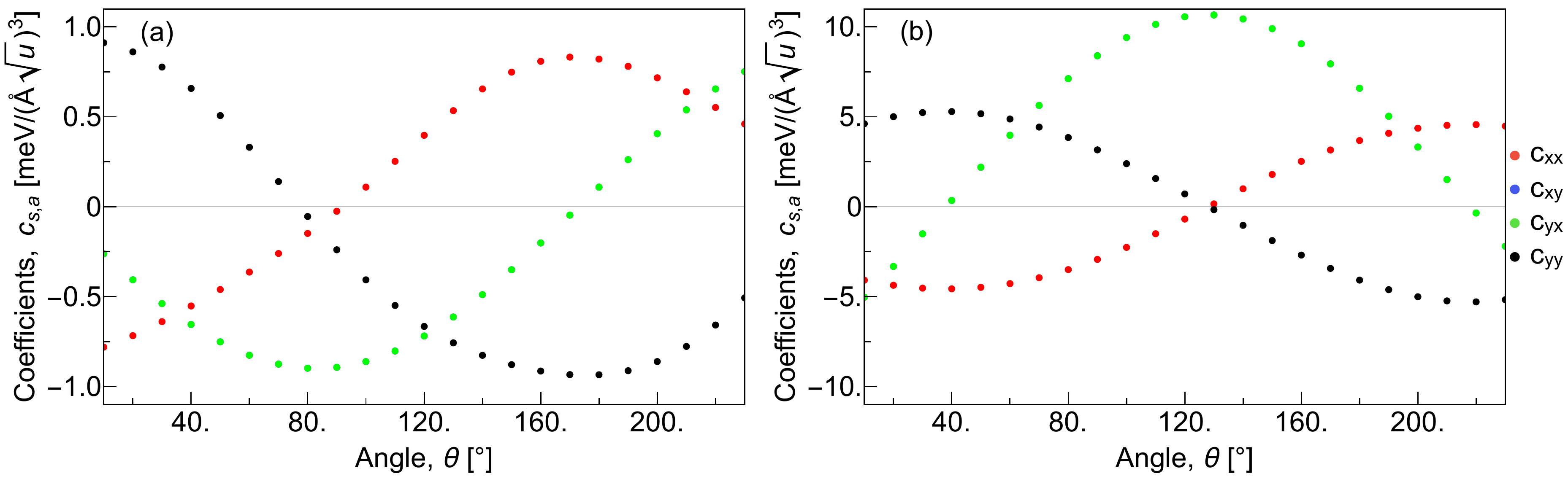}
\caption{Angle-dependent coupling-tensor coefficients for IRS for the (a) 40.45~THz mode and (b) 40.5~THz mode.}
\label{fig:coefficients-IRS}
\end{figure}

The interation potential for IRS is given by
\begin{equation}
V_\mathrm{int} =
\begin{pmatrix}
Q_{n,x} & Q_{n,y}
\end{pmatrix}
\underbrace{
\begin{pmatrix}
0 & c \\
c & 0
\end{pmatrix}
}_{\textstyle c_{s,a}}
\begin{pmatrix}
Q_{n,x} \\
Q_{n,y}
\end{pmatrix}
Q_{s,a},
\quad
V_\mathrm{int} =
\begin{pmatrix}
Q_{n,x} & Q_{n,y}
\end{pmatrix}
\underbrace{
\begin{pmatrix}
c & 0 \\
0 & -c
\end{pmatrix}
}_{\textstyle c_{s,b}}
\begin{pmatrix}
Q_{n,x} \\
Q_{n,y}
\end{pmatrix}
Q_{s,b}.
\end{equation}
The ionic Raman tensor $c_{s,a/b}$ is given by nonlinear three-phonon coupling, which we computed by displacing the high-frequency intralayer modes and the shear modes for amplitudes ranging from -0.8 to 0.8~$\text{\AA}\sqrt{u}$ and fitting the calculated potential energy surface to the above interaction potential. To probe the in-plane energy surface, we repeated this calculation for different angles of the shear-mode components ranging from $10\degree$ to $240\degree$, shown in Fig.~\ref{fig:coefficients-IRS}. We find the value of the independent tensor component $c$ to be 0.93~meV/(\AA$\sqrt{u}$)$^3$ for the 40.45~THz mode and 5.3~meV/(\AA$\sqrt{u}$)$^3$ for the 40.5~THz mode.


\subsection*{Impulsive stimulated Raman scattering and THz-sum frequency excitation}
\begin{figure}[h]
\centering
\includegraphics[width=17.5cm]{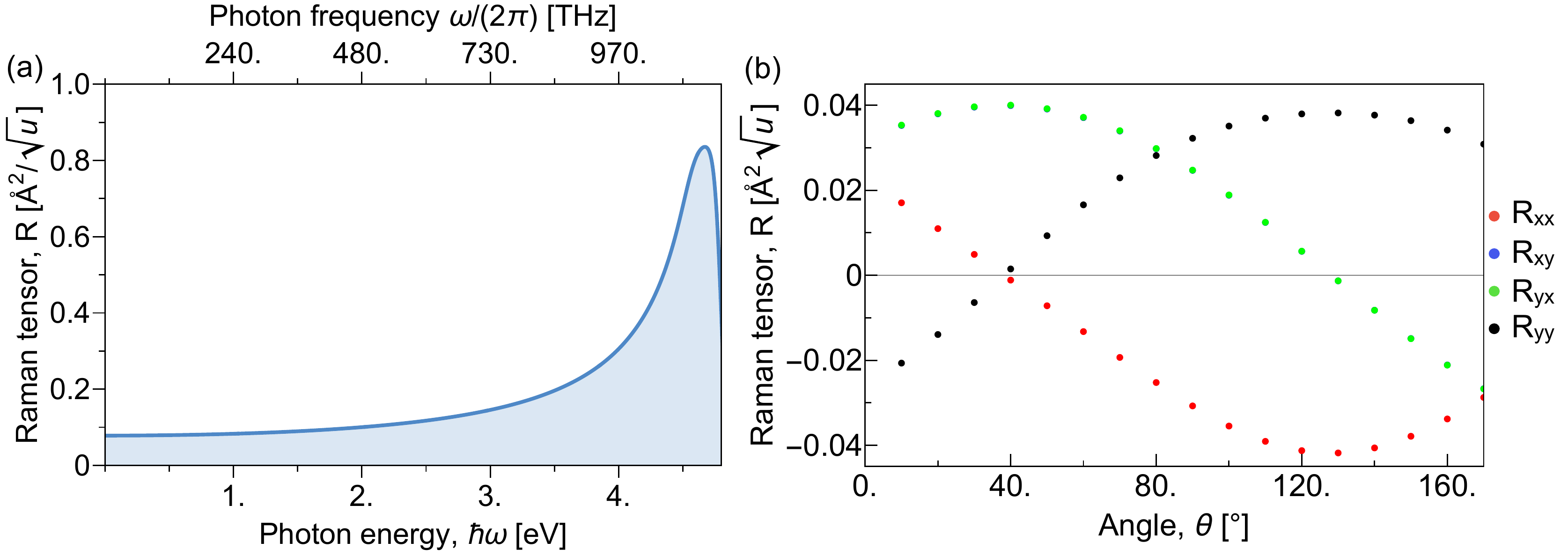}
\caption{(a) Frequency-dependent Raman-tensor coefficient $R$ for the shear modes. (b) Angle-dependent Raman-tensor coefficients.
}
\label{fig:RamanTensor}
\end{figure}

For both ISRS and THz-SFE, the interaction potential have the same symmetry and can be written as
\begin{equation}
V_\mathrm{int} =
\epsilon_{0}
\begin{pmatrix}
E_{x} & E_{y}
\end{pmatrix}
\underbrace{
\begin{pmatrix}
0 & R \\
R & 0
\end{pmatrix}
}_{\textstyle R_{s,a}}
\begin{pmatrix}
E_{x} \\
E_{y}
\end{pmatrix}
Q_{s,a},
\quad
V_\mathrm{int} =
\epsilon_{0}
\begin{pmatrix}
E_{x} & E_{y}
\end{pmatrix}
\underbrace{
\begin{pmatrix}
R & 0 \\
0 & -R
\end{pmatrix}
}_{\textstyle R_{s,b}}
\begin{pmatrix}
E_{x} \\
E_{y}
\end{pmatrix}
Q_{s,b}.
\end{equation}
The Raman tensor is given by $R_{s,a/b}=V_c \partial\epsilon/\partial Q_{s,a/b}$, where $V_c=36.43$~\AA$^3$ is the volume of the unit cell of the bilayer without vacuum and $\epsilon$ the frequency-dependent dielectric tensor. We computed the Raman tensor by evaluating $\epsilon$ for displacements of the shear modes for amplitudes ranging from -0.8 to 0.8~$\text{\AA}\sqrt{u}$, shown in Fig.~\ref{fig:RamanTensor}(a). To probe the in-plane energy surface, we repeated this calculation for different angles of the shear-mode components ranging from $10\degree$ to $170\degree$, shown in Fig.~\ref{fig:RamanTensor}(b). For the independent Raman-tensor component $R$, we obtained  $R(800~\text{nm})=0.09$~\AA$^2\sqrt{u}$ for ISRS and $R(0)=0.04$~\AA$^2\sqrt{u}$ for THz-SFE.


\subsection*{Infrared resonant Raman scattering}

\begin{figure}[t]
\centering
\includegraphics[width=15.5cm]{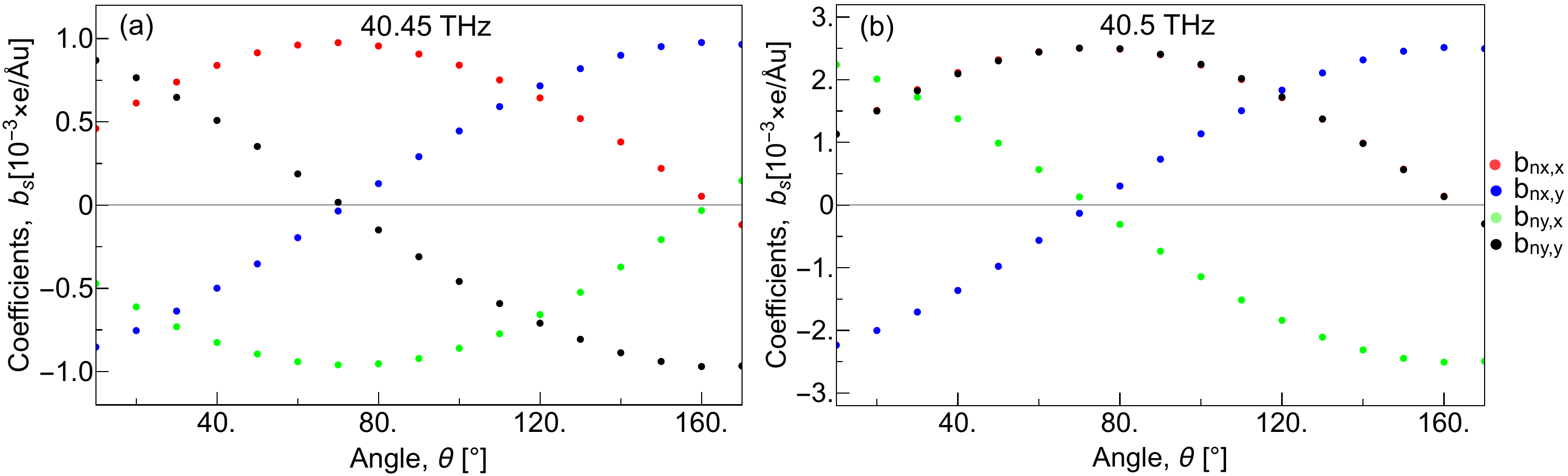}
\caption{Angle-dependent coupling-tensor coefficients for IRRS for the (a) 40.45~THz mode and (b) 40.5~THz mode.}
\label{fig:coefficients-IRRE}
\end{figure}

For IRRS, the interaction potential for the shear-mode components coupling to the electric field of the laser pulse and the high-frequency intralayer modes can be written as
\begin{equation}
V_\mathrm{int} =
\begin{pmatrix}
E_{x} & E_{y}
\end{pmatrix}
\underbrace{
\begin{pmatrix}
0 & b \\
b & 0
\end{pmatrix}
}_{\textstyle b_{s,a}}
\begin{pmatrix}
Q_{n,x} \\
Q_{n,y}
\end{pmatrix}
Q_{s,a}
\quad
V_\mathrm{int} =
\begin{pmatrix}
E_{x} & E_{y}
\end{pmatrix}
\underbrace{
\begin{pmatrix}
b & 0 \\
0 & -b
\end{pmatrix}
}_{\textstyle b_{s,b}}
\begin{pmatrix}
Q_{n,x} \\
Q_{n,y}
\end{pmatrix}
Q_{s,b}.
\end{equation}
The infrared resonant Raman tensor is given by $b_{s,a/b}=\partial \mathbf{Z}_{n}/\partial \mathbf{Q}_{s,a/b}$. We computed the tensor by displacing the shear modes for amplitudes ranging from -0.8 to 0.8~$\text{\AA}\sqrt{u}$ and then evaluating the mode effective charges of the high-frequency intralayer modes. To probe the in-plane energy surface, we repeat this calculation for different angles of the shear-mode components ranging from $10\degree$ to $170\degree$, shown in Fig.~\ref{fig:coefficients-IRRE}. We find the magnitude of the independent tensor component $b$ to be 0.001~$e/\text{\AA}u$ for the 40.45~THz mode and 0.0025~$e/\text{\AA}u$ for the 40.5~THz mode. To induce the scattering process, the high-frequency intralayer modes are excited in the same way as in IRS, yielding the same amplitude shown in Fig.~\ref{fig:modeIR}. 

%


\section*{Ferroelectric switching thresholds of the five mechanisms}
\begin{figure*}[h]
\centering
\includegraphics[width=15.5cm]{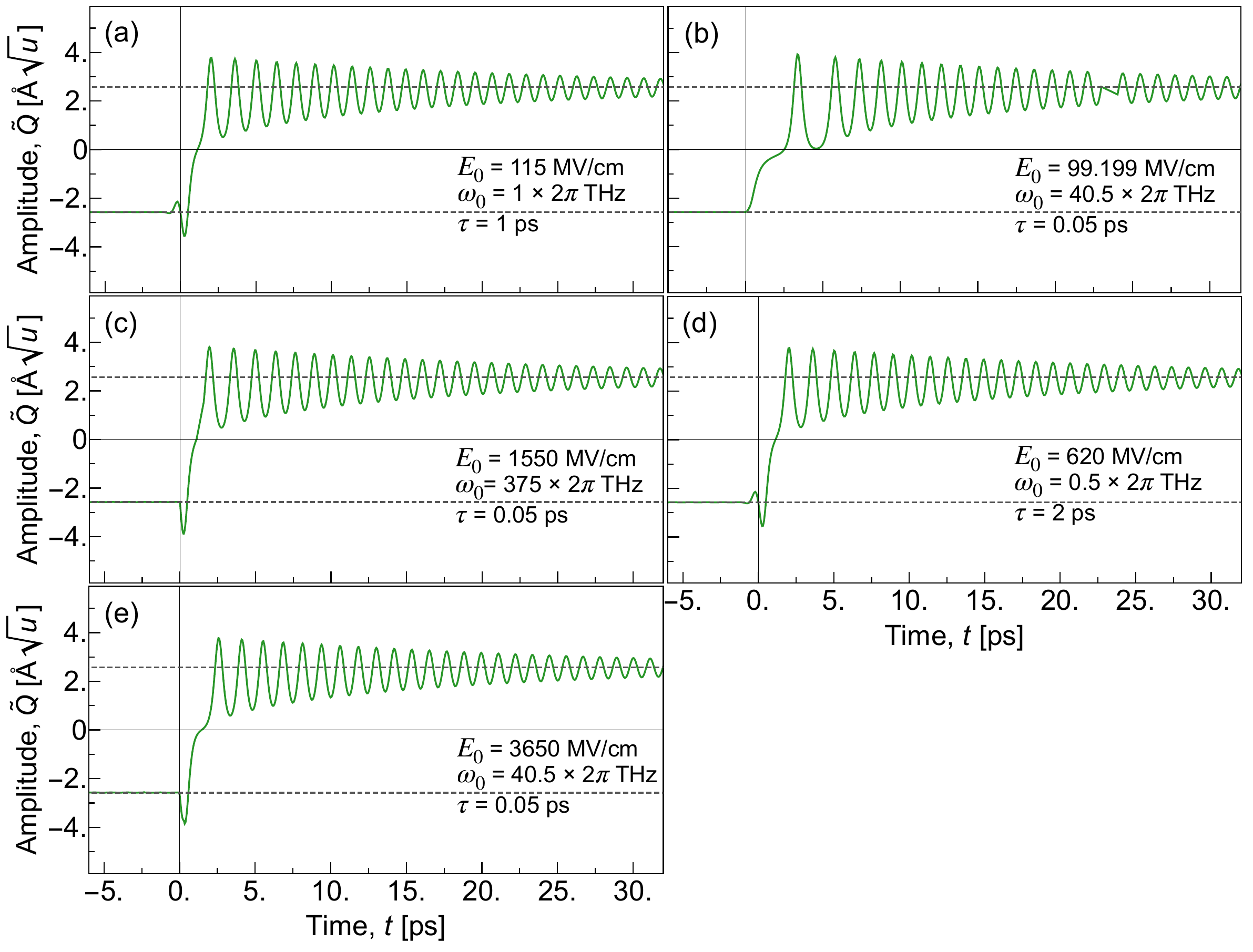}
\caption{Ferroelectric switching thresholds for the various mechanisms. Time evolution of the generalized shear-mode coordinate $\tilde{Q}$ for (a) IRA, (b) IRS, (c) ISRS, (d) THz-SFE, and (e) IRRS. Ferroelectric switching is shown for all five mechanism with the minimum values of peak electric fields $E_0$, pulse frequency $\omega_{0}$, and pulse duration $\tau$ indicated.
}
\label{fig:Switching_Allmechanism}
\end{figure*}

Here we discuss the ferroelectric switching threshold for all five excitation mechanisms. The potential energy of the double well is represented by the generalized shear-mode coordinate $\tilde{Q}$ and can be fitted accurately by a polynomial of the form $\tilde{V} = \alpha \tilde{Q}^{2}+\beta \tilde{Q}^{4}+\gamma \tilde{Q}^{6}+\delta\tilde{Q}^{8}$. The calculated parameters can be found in Table~\ref{tab:POtential-well-coefficients}. The dynamics of the shear-mode within the double-well potential are captured by the equation of motion
\begin{align}
\ddot{\tilde{Q}}+\kappa \dot{\tilde{Q}}+\frac{\partial \tilde{V}}{\partial \tilde{Q}}+\frac{\partial {V_\mathrm{int}}}{\partial \tilde{Q}}= 0. 
\end{align}
%
%
\begin{table}[h]
    \centering
    \caption{Values of fitting parameter of double-well potential. $n$ denotes the order of phonon amplitude.}
    \begin{tabular}{lllll}
    \hline\hline
       Coefficients  & $\alpha$  &$\beta$  &$\gamma$  &$\delta$\\
    \hline
     Values in meV/($\text{\AA}\sqrt{u}$)$^n$   &-1.1   &0.1 &$-0.19\times 10^{-6}$  &$0.12\times10^{-8}$ \\
    \hline\hline
    \end{tabular}
    \label{tab:POtential-well-coefficients}
\end{table}
In Fig.~\ref{fig:Switching_Allmechanism}, we show the switching dynamics for all five mechanisms. The required values for $E_0$ and $\tau$ indicate that all mechanisms require unfeasibly large pulse energies to overcome the energy barrier.


 \begin{figure}[b]
 \centering
\includegraphics[width=9cm]{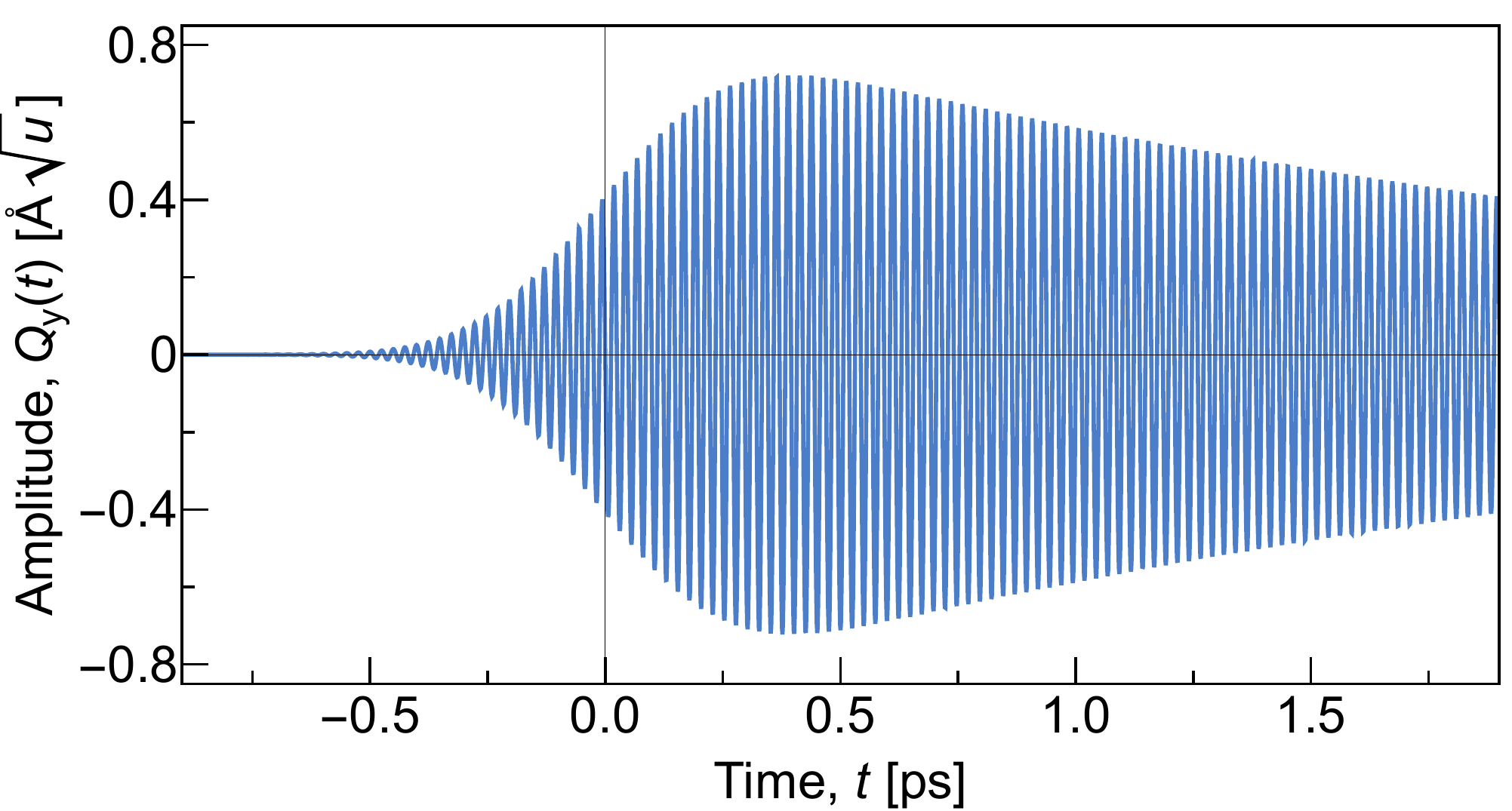}[b]
 \caption{Time evolution of amplitude of the 40.5~THz $E$ mode following resonant excitation by the laser pulse for a peak electric field $E_0= 20$~MV/cm and pulse duration $\tau=0.5$~ps.
 }
\label{fig:mode_12}
 \end{figure}
 
\section*{Time-dependent double-well potential and ferroelectric switching}

\begin{figure*}[b]
\centering
\includegraphics[width=14cm]{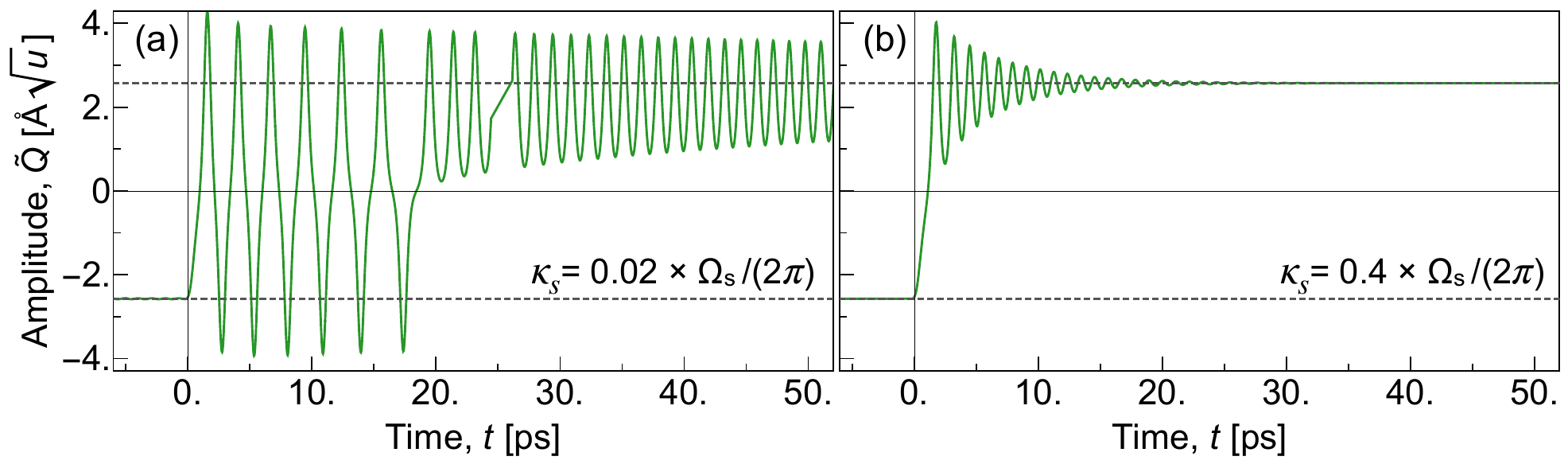}
\caption{Time evolution of ferroelectric switching for (a) low and (b) high shear-mode damping, initiated by a pulse with a peak electric field of $E_{0}=21$~MV/cm and a pulse duration of $\tau= 0.05$~ps.}
\label{fig:Switching_comparison}
\end{figure*}

The shear-mode potential is modified by the displacement of the atoms along the eigenvectors of the high-frequency intralayer $E$ mode at 40.5~THz. The potential becomes a functional of the time-dependent amplitude of the 40.5~THz mode with modified coefficients for $\tilde{V}$. We display the fitted time-dependent potential below and the values of the fitted coefficients in Table~\ref{tab:tilt POtential-well-coefficients}. 
\begin{align}\label{eq:tilting}
    \tilde{V}[Q_2(t)]= & (a_1 + a_2{Q}_2 + 
   a_3{Q}_2^2)+ (b_2{Q}_2+ 
   b_3{Q}_2^2)\tilde{Q}
    +(\alpha + c_2{Q}_2+ c_3{Q}_2^2)\tilde{Q}^2 +  (d_2{Q}_2+d_3{Q}_2^2)\tilde{Q}^3 
   + (\beta + e_2{Q_2} + e_3{Q_2^2})\tilde{Q}^4 \nonumber\\
   & + (f_2{Q}_2+f_3{Q}_2^2)\tilde{Q}^5 
   +(\gamma + g_2{Q}_2 + g_3{Q}_2^2)\tilde{Q}^6 + (h_2{Q}_2+h_3{Q}_2^2)\tilde{Q}^7 
  + (\delta + i_2{Q}_2+ i_3{Q}_2^2)\tilde{Q}^8
\end{align}
Figure~\ref{fig:mode_12} shows the time evolution of the 40.5~THz mode under tilting switching in the main text. Figure~\ref{fig:Switching_comparison} shows the tilting switching dynamics for low and high shear-mode dampings.


\begin{table}[h]
    \centering
    \caption{Values of fitting parameters for the time-dependent double-well potential. $n$ denotes the order of phonon amplitude.}
    \begin{tabular}{ll|ll}
    \hline\hline
       Coefficients  & Values  &Coefficients  & Values  \\
       meV/(\AA$\sqrt{u}$)$^n$& &meV/(\AA$\sqrt{u}$)$^n$ &\\
    \hline
     $a_1$   &3.6              &$e_3$   &-$6.6\times10^{-2}$ \\
     $a_2$   & -15.7             &$f_2$   &-$9.2\times10^{-3}$ \\
     $a_3$   & -16.8             &$f_3$   &-$1.3\times10^{-3}$ \\
     $b_2$   &-8               &$g_2$   &-$6.6\times10^{-5}$ \\
     $b_3$   &-5.2              &$g_3$   &$1.2\times10^{-3}$  \\
     $c_2$   & $5.3\times10^{-2}$ &$h_2$   &$7.3\times10^{-5}$ \\
     $c_3$   & 1                &$h_3$  & $9.6\times10^{-6}$\\
     $d_2$   &  0.31             &$i_2$  &$1\times10^{-6}$   \\
     $d_3$   & $5.5\times10^{-2}$  &$i_3$  & -$7.2\times10^{-6}$\\
     $e_2$   & $8.8\times10^{-4}$   &       & \\
    \hline\hline
    \end{tabular}
    \label{tab:tilt POtential-well-coefficients}
\end{table}


\end{document}